\documentclass[aps,prd,twocolumn,showpacs,nofootinbib]{revtex4}

\usepackage{amsfonts}
\usepackage{mathrsfs}
\usepackage{graphicx}
\usepackage{amsmath}
\usepackage{amssymb}
\usepackage{color}

\usepackage[colorlinks=true,linkcolor=blue,citecolor=blue, urlcolor=blue]{hyperref}

\begin{document}

\title{Degeneracy Relations in QCD and the Equivalence of Two Systematic All-Orders Methods for Setting the Renormalization Scale}

\author{Huan-Yu Bi}
\email{bihy@cqu.edu.cn}
\author{Xing-Gang Wu}
\email{wuxg@cqu.edu.cn}
\author{Yang Ma}
\email{mayangluon@cqu.edu.cn}
\author{Hong-Hao Ma}
\email{mahonghao@cqu.edu.cn}
\address{Department of Physics, Chongqing University, Chongqing 401331, P.R. China}
\address{Institute of Theoretical Physics, Chongqing University, Chongqing 401331, P.R. China}

\author{Stanley J. Brodsky}
\email{sjbth@slac.stanford.edu}
\affiliation{SLAC National Accelerator Laboratory, Stanford University, Stanford, California 94039, USA}

\author{Matin Mojaza}
\email{mojaza@nordita.org}
\affiliation{Nordita, KTH Royal Institute of Technology and Stockholm University, Roslagstullsbacken 23, SE-10691 Stockholm, Sweden}

\date{\today}

\begin{abstract}

The Principle of Maximum Conformality (PMC) eliminates QCD renormalization scale-setting uncertainties using fundamental renormalization group methods. The resulting scale-fixed pQCD predictions are independent of the choice of renormalization scheme and show rapid convergence. The coefficients of the scale-fixed couplings are identical to the corresponding conformal series with zero $\beta$-function. Two all-orders methods for systematically implementing the PMC-scale setting procedure for existing high order calculations are discussed in this article. One implementation is based on the PMC-BLM correspondence \mbox{(PMC-I)}; the other, more recent, method \mbox{(PMC-II)} uses the ${\cal R}_\delta$-scheme, a systematic generalization of the minimal subtraction renormalization scheme. Both approaches satisfy all of the principles of the renormalization group and lead to scale-fixed and scheme-independent predictions at each finite order. In this work, we show that PMC-I and PMC-II scale-setting methods are in practice equivalent to each other. We illustrate this equivalence for the four-loop calculations of the annihilation ratio $R_{e^+ e^-}$ and the Higgs partial width $\Gamma(H\to b\bar{b})$. Both methods lead to the same resummed (`conformal') series up to all orders. The small scale differences between the two approaches are reduced as additional renormalization group $\{\beta_i\}$-terms in the pQCD expansion are taken into account. We also show that {\it special degeneracy relations}, which underly the equivalence of the two PMC approaches and the resulting conformal features of the pQCD series, are in fact general properties of non-Abelian gauge theory.

\end{abstract}

\pacs{12.38.Bx, 12.38.Aw, 11.15.Bt}

\maketitle

A primary problem for perturbative QCD (pQCD) analyses is how to systematically set the renormalization scale of the QCD running coupling to achieve precise fixed-order predictions for physical observables~\cite{Wu:2013ei}.  A valid prediction at any finite order should be independent of the choice of the renormalization scheme, since the choice of the scheme is a theoretical convention. The ``Principle of Maximum Conformality" (PMC)~\cite{Brodsky:2011ta, Brodsky:2011ig, Mojaza:2012mf, Brodsky:2013vpa} eliminates the renormalization scheme- and scale- ambiguities order by order in perturbation theory, consistent with fundamental renormalization group methods~\cite{Stueckelberg:1953dz, Peterman:1978tb, GellMann:1954fq, rge4}; it is also the principle underlying the well-known Brodsky-Lepage-Mackenzie (BLM) approach~\cite{Brodsky:1982gc}. In the Abelian limit, the PMC is the standard scale-setting method used for precision tests of quantum electrodynamics (QED). The elimination of the renormalization scale ambiguity removes an unnecessary systematic error for pQCD predictions, thus providing scheme-independent precision tests of the Standard Model and improving the sensitivity to new physics.

The renormalization group equation (RG-equation) determines the running of the gauge coupling from the analytic properties of the $\beta$-function:
\begin{equation}
\beta(a_s(\mu))=\mu^2\frac{{\rm d} a_s(\mu)} {{\rm d}\mu^2} = -a_s^2(\mu) \sum_{i=0}^\infty \beta_i a^{i}_s(\mu), \label{rge}
\end{equation}
where a perturbative expansion of the $\beta$-function in terms of the coupling $a_s=\alpha_s/4\pi$ is assumed. The expressions for $\beta_{0}, \ldots, \beta_3$ in the modified minimal-subtraction scheme ($\overline{\rm MS}$-scheme) can be found in Refs.~\cite{Politzer:1973fx, Tarasov:1980au, vanRitbergen:1997va, Chetyrkin:2004mf, Czakon:2004bu}. One can then use a Taylor-expansion to derive a scale-displacement relation for the running coupling at two different scales $\mu_1$ and $\mu_2$:
\begin{equation}
a_s(\mu_2) = a_s(\mu_1) + \sum_{n=1}^\infty \frac{1}{n!} \left. \frac{{\rm d}^n a_s(\mu)}{({\rm d} \ln\mu^2)^n}\right|_{\mu=\mu_1} \left(\ln\frac{\mu_2^2}{\mu_1^2}\right)^n . \label{running}
\end{equation}

The RG-equation and the asymptotically expanded $\beta$-function in \eqref{rge} can be used to recursively establish a perturbative $\beta$-pattern at each order:
\begin{eqnarray}
a_s(\mu_2) &=& a_s(\mu_1)- \beta_{0} \ln\left(\frac{\mu_2^{2}} {\mu_1^2}\right) a^{2}_s(\mu_1) \nonumber\\
&& +\left[\beta^2_{0} \ln^2 \left(\frac{\mu_2^{2}}{\mu_1^2}\right) -\beta_{1} \ln \left(\frac{\mu_2^{2}} {\mu_1^2}\right)\right] a^{3}_s(\mu_1) \nonumber\\
&& + \left[ \frac{5}{2}{\beta_0}{\beta_1} \ln^2 \left(\frac{\mu_2^{2}}{\mu_1^2}\right) - \beta_0^3\ln^3 \left(\frac{\mu_2^{2}}{\mu_1^2}\right) \right. \nonumber\\
&& \left. \quad\quad - \beta_2 \ln \left(\frac{\mu_2^{2}}{\mu_1^2}\right) \right] a^{4}_s(\mu_1) + \ldots .  \label{scaledis}
\end{eqnarray}
The PMC utilizes this perturbative $\beta$-pattern to systematically set the scales of the running coupling at each order in a pQCD expansion; the coefficients of the resulting series thus match the coefficients of the corresponding conformal theory with $\beta=0$. Thus the divergent $\alpha_s^n \beta_i^n n!$ ``renormalon" contributions are absorbed into the scale-fixed running couplings. The pQCD convergence of the resummed series is generally improved due to the elimination of those renormalon terms. This is the same principle used in QED where all $\beta$-terms resulting from the vacuum polarization corrections to the photon propagator are absorbed into the scale of the running coupling. As in QED, the scales are physical in the sense that they reflect the virtuality of the gluon propagators at a given order, as well as setting the effective number $n_f$ of active flavors. The resulting resummed pQCD expression thus determines the relevant ``physical" scales for any physical observable, thereby providing a solution to the renormalization scale-setting problem. There can be a small residual scale-uncertainty in the final expression due to the truncation of the $\beta$-function; however, these residual uncertainties are highly suppressed even for lower-order predictions.

The scale-displacement relation \eqref{running} provides the simplest example of the PMC: If one resums all-orders $\{\beta_i\}$-terms in the right-hand-side of Eq.\eqref{scaledis}, one (trivially) obtains $a_s(\mu_2)\equiv a_s(\mu_2)|_{\rm PMC}$ at any fixed order, independent of $\mu_1$ and the choice of scheme; this agrees with the reflexivity property of the renormalization group~\cite{Brodsky:2012ms}.

In this paper we will show how to systematically implement the PMC scale-setting if one starts with an existing high-order pQCD calculations for a physical observable. Such calculations which are available in the literature are usually performed in a conventional renormalization scheme such as the $\overline{\rm MS}$-scheme, assuming a single initial scale $\mu$. Two all-orders approaches for implementing the PMC procedure for existing high-order calculations have been suggested. One is based on the PMC-BLM correspondence \mbox{(PMC-I)}~\cite{Brodsky:2011ta}; the other method \mbox{(PMC-II)}~\cite{Mojaza:2012mf, Brodsky:2013vpa} utilizes the pattern of $\beta$ terms illuminated by the ${\cal R}_\delta$-scheme, which was introduced in Ref.\cite{Mojaza:2012mf} as a generalization of the MS-like renormalization schemes, thus enabling one to obtain nontrivial information on the pQCD series. Both approaches satisfy all of the principles of the renormalization group~\cite{Brodsky:2012ms}, and they lead to scale-fixed and scheme-independent predictions at any finite order~\cite{Brodsky:2011ta, Brodsky:2011ig, Mojaza:2012mf, Brodsky:2013vpa}. Both implementations have been successfully used for making scale-fixed and rapidly converging predictions for a number of high-order high-energy processes~\cite{Brodsky:2012sz, Brodsky:2012ik, Brodsky:2012rj, Wang:2014sua, Wang:2014aqa, Wang:2014wua, Wang:2013akk, Wang:2013bla, Wu:2014iba, Shen:2015cta}. We shall show that the two PMC methods are, in practice, equivalent. Any small residual scale difference between the two approaches is systematically reduced as more RG $\{\beta_i\}$-terms in the perturbative QCD expansion are known. Both implementations of the PMC allow one to determine the PMC scales of a process order-by-order in pQCD.

In general, a pQCD prediction for a physical observable can be written as an expansion~\footnote{We do not consider quark mass renormalization in this paper and will assume $n\geq1$ in our discussions. For the case of $n=0$, i.e. when the tree-level result does not involve strong interactions, one can implement the PMC from the second $a^{1}$-term and all the following formulas apply.}
\begin{equation}
\rho = \sum^{m}_{i=1} \left(\sum^{i-1}_{j=0} c_{i,j} n_f^{j}\right) a_s^{n+i-1}(\mu) +\ldots, \label{nf}
\end{equation}
where the symbol ``$\ldots$" means even higher-order contributions, $\mu$ is the initial scale, The $n_f$-terms count the number of active flavors which arise from light-quark loop contributions to the $\beta$-function. All of the explicitly written $n_f$-terms thus pertain to the RG $\{\beta_i\}$-terms, and -- as in the BLM procedure -- they provide a guide to setting the renormalization scales. It is important to note that the coefficients $c_{i,j}$ may also contain $n_f$-terms which are ultraviolet finite and are not associated with the $\beta$-function. For example, terms arising from the ultraviolet finite quark-loop contributions to the three-gluon and four-gluon vertices will be kept unchanged during PMC scale-setting.

After applying PMC-I and PMC-II, all of the $n_f$-terms which are governed by the RG-equation will be resummed into the running coupling; the pQCD series in Eq.~\eqref{nf} then changes to the resummed `conformal' series:
\begin{eqnarray}
\rho &=& \sum^{m}_{i=1} r_{i,0}^{\rm M} {a_s^{n+i-1}}(Q_{{\rm M},i})+ \ldots , \label{PMCseries}
\end{eqnarray}
where ${\rm M}$=I or II for PMC-I or PMC-II, respectively. The coefficients $r^{\rm M}_{i,0}$ are the $\beta$-independent `conformal' coefficients and $Q_{{\rm M},i}$ are the PMC scales, where $i=1$ stands for the leading-order (LO) one, $i=2$ stands for the next-to-leading (NLO) on, etc.  The resulting coefficients are conformal in the sense that they are UV finite, scheme-independent, and do not depend on the resummation schemes, PMC-I, PMC-II, or any other implementation following PMC. This will be shown explicitly for PMC-I and \mbox{PMC-II}.

Let us describe the different implementations of PMC in more detail:

PMC-I allows one to obtain the correct PMC scales using a step-by-step method without first transforming the $n_f$-terms into the $\{\beta_i\}$-terms. This procedure is based on the observation that one can rearrange all the Feynman diagrams of a process in form of a cascade; i.e., the ``new" terms emerging at each order can be equivalently regarded as a one-loop correction to all the ``old" lower-order terms. All of the $n_f$-terms can then be absorbed into the running coupling following the basic $\beta$-pattern in the scale-displacement formula, i.e. Eq.\eqref{scaledis}. The resulting PMC-I scales are themselves expressed as perturbative expansions with the same $\beta$-pattern of Eq.(\ref{scaledis}). More explicitly, the PMC-I scales can be derived in the following way: The LO PMC-I scale $Q_{{\rm I},1}$ is obtained by eliminating all the $n_f$-terms with the highest power at each order, and at this step, the coefficients of the lower-power $n_f$-terms are changed simultaneously to ensure that the correct LO $\alpha_s$-running is obtained; the NLO PMC scale $Q_{{\rm I},2}$ is obtained by eliminating the $n_f$-terms of one less power in the new series obtaining a third series with less $n_f$-terms; and so on until all $n_f$-terms are eliminated. The step-by-step coefficients for the $n_f$-series can be found in Ref.\cite{Brodsky:2011ta}. After performing these order-by-order scale shifts $\mu\to Q_{{\rm I},1}$, $Q_{{\rm I},1} \to Q_{{\rm I},2}$, $Q_{{\rm I},2} \to Q_{{\rm I},3}$, $\cdots$, one eliminates all the $n_f$-terms associated with the $\alpha_s$-running and derive the conformal series.

It is noted that the PMC scale-setting can be automatically implemented in a higher order pQCD calculation if one uses the ${\cal R}_\delta$-scheme. The usual subtraction constant $\ln4\pi - \gamma_E $ is generalized to $\ln4\pi - \gamma_E -\delta$. The resulting dependence on the extra constant $\delta$ flags all of the terms in the pQCD series proportional to the $\beta$-function~\cite{Mojaza:2012mf, Brodsky:2013vpa}. Thus, in contrast to PMC-I, PMC-II first transforms the $n_f$-series at each order into the specific $\beta$-pattern dictated by the ${\cal R}_\delta$-scheme; the resulting $\beta$-pattern leads directly to the PMC scales and the conformal series. In this sense, PMC-II is a theoretical improvement of PMC-I, since PMC-I only determines the overall effective scales without determining the perturbative terms which lead to those values. Due to the fact that the running of $\alpha_s$ at each order has its own $\{\beta_i\}$-series as governed by the RG-equation, the $\beta$-pattern for the pQCD series at each order is a superposition of all of the $\{\beta_i\}$-terms which govern the evolution of the lower-order $\alpha_s$ contributions at this particular order. The resulting $\beta$-pattern is then in general different from the $\beta$-pattern of Eq.(\ref{scaledis}).

PMC-II suggestes that the coefficients of the $\{\beta_i\}$-terms in the $\beta$-pattern can be fixed by requiring a ``degeneracy relation" among different $\{\beta_i\}$-terms at different orders; the result resembles a skeleton-like expansion~\cite{Brodsky:2013vpa, Mojaza:2012mf}. By resumming the $\{\beta_i\}$-series according to this expansion, one also correctly reproduces the Abelian $N_c \to 0$ limit of the observables~\cite{Brodsky:1997jk}. Thus one can simultaneously determine the PMC scales $Q_{{\rm II},i}$ at all orders from their initial values $\mu$; i.e. $\mu\to Q_{{\rm II},1}$, $\mu\to Q_{{\rm II},2}$, $\mu\to Q_{{\rm II},3}$, $\cdots$, by resumming the $\{\beta_i\}$-terms into the running couplings in the skeleton-like form.

The degeneracy relations introduced by PMC-II were originally obtained by studying the pQCD series in the ${\cal R}_\delta$-scheme~\cite{Mojaza:2012mf}. Let us show that theses relations are required by the conformality of the final series. Using the RG-equation and the scale displacement relation, we can write down the most general $\beta$-pattern for the pQCD approximant of a physical observable. For example, Eq.(\ref{nf}) can be rewritten as
\begin{widetext}
\begin{eqnarray}
\rho &=& {r^{\rm II}_{1,0}}{a_s^n}(\mu) + \left({r^{\rm II}_{2,0}}+{n}{\beta_0}{r^{\rm II}_{2,1}} \right) {a_s^{n + 1}}{(\mu)}+ \left({r^{\rm II}_{3,0}}+ {n}{\beta_1}{r^{\rm II*}_{2,1}}+ (n + 1){\beta_0}{r^{\rm II}_{3,1}}+\frac{{n^2 + n}}{{2}}\beta_0^2{r^{\rm II}_{3,2}}\right){a_s^{n + 2}}(\mu) \nonumber\\
&& + \left({r^{\rm II}_{4,0}} + {n} {\beta_2}{r^{\rm II**}_{2,1}}+ ({{n + 1}} ){\beta_1}{r^{\rm II*}_{3,1}}+ (n + 2){\beta_0}{r^{\rm II}_{4,1}}+ \frac{2n^2 + 3n}{2}{\beta_1}{\beta_0}{r^{\rm II*}_{3,2}}+ \frac{n^2+ 3n + 2}{2}\beta_0^2{r^{\rm II}_{4,2}} \right. \nonumber\\
&& \left. +\frac{n^3 + 3n^2 + 2n}{6}\beta_0^3{r^{\rm II}_{4,3}} \right) a_s^{n +3}(\mu) + \cdots . \label{PMCIIs}
\end{eqnarray}
By applying the PMC-II procedures~\cite{Mojaza:2012mf,Brodsky:2013vpa}, we obtain
\begin{eqnarray}
\rho &=& r_{1,0}^{{\rm{II}}}{a_s^n}(Q_{{\rm II},1}) + r_{2,0}^{{\rm{II}}} \cdot {a_s^{n + 1}}(Q_{{\rm II},2}) + r_{3,0}^{{\rm{II}}} \cdot {a_s^{n + 2}}(Q_{{\rm II},3}) + r_{4,0}^{{\rm{II}}} \cdot {a_s^{n + 3}}(Q_{{\rm II},3}) +{n}\left[{r_{2,1}^{\rm II*}}- {r^{\rm II}_{2,1}}\right]{\beta _1}{a_s^{n + 2}}(\mu)\nonumber\\
&& +\left({n}{\beta _2}\left [{r^{\rm II**}_{2,1}}-{r^{\rm II}_{2,1}} \right ]+({{n + 1}}){\beta _1}\left [{r^{\rm II*}_{3,1}}-{r^{\rm II}_{3,1}}\right ]+\frac{{2n^2 + 3n}}{{2}}{\beta _1}{\beta _0}\left [{r^{\rm II*}_{3,2}}-{r^{\rm II}_{3,2}}\right ]\right) {a_s^{n + 3}}{(\mu)} + \cdots. \label{nod}
\end{eqnarray}
\end{widetext}
This step does not require the degeneracy relations. However, in order to ensure that $\rho$ is conformal; i.e., to not contain any $\{\beta_i\}$-terms, we get the required degeneracy relations: $r^{\rm II**}_{2,1}=r^{\rm II*}_{2,1}=r^{\rm II}_{2,1}$, $r^{\rm II*}_{3,1}=r^{\rm II}_{3,1}$, $r^{\rm II*}_{3,2}=r^{\rm II}_{3,2}$, etc.. Those degeneracy relations also ensure the elimination of all the remaining $a_s(\mu)$-terms in the pQCD series, we then obtain the required scale-fixed PMC prediction.

Alternatively, if one implements PMC-I and requires the uniqueness of the conformal coefficients for the two methods, we again find the degeneracy relations. To show this, we transform the general $\{\beta_i\}$-series (\ref{PMCIIs}) back to $n_f$-series, and by further applying PMC-I, one finds
\begin{eqnarray}
r_{1,0}^{\rm I}&=& r^{\rm II}_{1,0}, \nonumber\\
r_{2,0}^{\rm I}&=& r^{\rm II}_{2,0},\; \nonumber\\
r_{3,0}^{\rm I}&=& r^{\rm II}_{3,0} + 7 nC_A^2(r^{\rm II}_{2,1} - r^{\rm II*}_{2,1}) + 11 n C_A C_F (r^{\rm II}_{2,1} - r^{\rm II*}_{2,1}) \nonumber \\
r_{4,0}^{\rm I}&=& r^{\rm II}_{4,0} -\frac{1}{3} C_A^2 (n (151 r^{\rm II}_{2,1} -228 r^{\rm II*}_{2,1} +77 r^{\rm II**}_{2,1}) C_F \nonumber\\
&& -21 (n+1) (r^{\rm II}_{3,1}-r^{\rm II*}_{3,1} ) )+ \frac{7}{24} n (41 r^{\rm II}_{2,1} +120 r^{\rm II*}_{2,1} \nonumber \\
&& -161 r^{\rm II**}_{2,1} ) C_A^3 -\frac{11}{2} C_A C_F (n (7 r^{\rm II}_{2,1}-6 r^{\rm II*}_{2,1}\nonumber\\
&& -r^{\rm II**}_{2,1} ) C_F-2 (n+1) (r^{\rm II}_{3,1}-r^{\rm II*}_{3,1} ) ). \nonumber
\end{eqnarray}
Thus to ensure the conformality of the \mbox{PMC-I} and the \mbox{PMC-II} final expressions; i.e. $r_{i,0}^{\rm I} \equiv r_{i,0}^{\rm II}$, we are immediately led to the degeneracy relations. Following the same procedures, we can demonstrate the equivalence of conformal series up to any order. This equivalence can be explained by the fact that the scale-displacement relation (\ref{running}) acts only on the purely non-conformal $\{\beta_i\}$-series, and PMC-I and PMC-II only differ in eliminating the $n_f$-terms -- either by using the RG $\beta$-pattern directly, or by using the super-positioned RG $\beta$-pattern. Thus, if one transforms the PMC-I prediction to the one of PMC-II, or vice versa, the conformal coefficients are not altered. More explicitly, by using the $\beta$-function to the highest known order; i.e., four-loops, the conformal coefficients for any semi-simple Lie gauge group with $n_f$-fermions and $N_c$-colors are:
\begin{eqnarray}
r_{1,0}^{\rm I, II}&=& c_{1,0} , \\
r_{2,0}^{\rm I, II}&=& c_{2,0} + \frac{{11{C_A}}}{{4{T_F}}}{c_{2,1}} , \\
r_{3,0}^{\rm I, II}&=&\frac{1}{{16T_F^2}} \left[ - (84C_A^2{T_F} + 132{C_A}{C_F}{T_F}){c_{2,1}} + \right. \nonumber\\
&& \left. 16T_F^2{c_{3,0}} + 44{C_A}{T_F}{c_{3,1}} + 121C_A^2{c_{3,2}}\right] ,\\
r_{4,0}^{\rm I, II}&=& \frac{1}{{64T_F^3}} [ 2{C_A}T_F^2 ( - 287C_A^2 + 1208{C_A}{C_F} + \nonumber\\
&& 924C_F^2 ) {c_{2,1}} - 48{C_A}T_F^2 ( {7{C_A} + 11{C_F}}){c_{3,1}} - \nonumber\\
&& 264C_A^2{T_F}\left( {7{C_A} + 11{C_F}} \right){c_{3,2}} + 64T_F^3{c_{4,0}}+ \nonumber\\
&& 176{C_A}T_F^2{c_{4,1}} + 484C_A^2{T_F}{c_{4,2}} + 1331C_A^3{c_{4,3}}],
\nonumber \\
\end{eqnarray}
where $C_A$, $C_F$ and $T_F$ are quadratic Casimir invariants~\cite{vanRitbergen:1998pn}. For a $SU(N_c)$-color group, we have $C_A=N_c$, $C_F=(N^2_c -1) / 2 N_c$ and $T_F=1/2$.

Since the non-conformal $\{\beta_i\}$-terms are eliminated in different ways, the PMC-I and PMC-II scales can in principle be different. Since they are both based on the RG-equation, the accuracy of the PMC scales depend heavily on how well we know the $\{\beta_i\}$-terms of the process. One observes that the PMC-I and PMC-II scales are themselves expressed as perturbative series, and their logarithmic differences will be suppressed by specific powers of $\alpha_s$. To quantify this, we define the logarithmic difference of the PMC-I and PMC-II scales by $\Delta_{i} =\ln {Q_{{\rm I},i}}/{Q_{{\rm II},i}}$, and the first three ones for a four-loop prediction read:
\begin{eqnarray}
\Delta_{1} &=&-\frac{3 \beta _1 f}{64 n^2 (n+1) c_{1,0}^2 c_{2,1} T_F^2}a_s(\mu)^2 + {\cal O}(a^3_s), \label{scale1} \\
\Delta_{2} &=&\frac{3 \beta_0 \left(5 C_A+3 C_F\right) f}{16 n (n+1)^2 c_{1,0} c_{2,1} T_F \left(11 c_{2,1} C_A+4 c_{2,0} T_F\right)} a_s(\mu) \nonumber\\
&& + {\cal O}(a^2_s) , \label{scale2} \\
\Delta_{3} &=& 3 C_A \left(7 C_A+11 C_F\right) \frac{ f}{g}+ {\cal O}(a_s), \label{scale3}
\end{eqnarray}
where
\begin{eqnarray}
f &=& 6 n^2 (n+1) c_{1,0}^2 c_{4,3} + \left(2 n^3+8 n^2+13 n+7\right) c_{2,1}^3 \nonumber\\
&& -6 n \left(n^2+3 n+3\right) c_{1,0} c_{3,2} c_{2,1}, \nonumber\\
g &=& 4 n (n+1) (n+2) c_{1,0} c_{2,1} (C_A^2 (84 c_{2,1} T_F-121 c_{3,2}) \nonumber\\
&& +44 C_A T_F (3 c_{2,1} C_F-c_{3,1})-16 c_{3,0} T_F^2). \nonumber
\end{eqnarray}
As indicated by Eqs.(\ref{scale1},\ref{scale2},\ref{scale3}), the LO logarithmic scale difference $\Delta_1$ starts at the order $a_s^2$, which changes to order $a^1_s$ for the NLO $\Delta_2$, and to order $a^0_s$ for the NNLO $\Delta_3$. The value of $\Delta_i$ can be qualitatively understood through the scale-displacement equation \eqref{scaledis}, i.e.,
\begin{displaymath}
a_s^i(Q_{{\rm II},i})-a_s^i(Q_{{\rm I},i}) = 2 i \beta_0 \Delta_i a_s^{i+1}(Q_{{\rm II},i}) +{\cal O}(a_s^{i+2}).
\end{displaymath}
The leading term of $\Delta_i$ may be of order $a_s^0$. Since the NLO $n_f$-term is uniquely fixed, the order $a_s^0$ of $\ln {Q_{{\rm I},1}}$ and $\ln {Q_{{\rm II},1}}$ must be equal. This explains why PMC-I, PMC-II, and also BLM, are exactly the same at the NLO level~\cite{Ma:2015dxa}. We further note that the order $a_s^1$ of $\ln {Q_{{\rm I},1}}$ and $\ln {Q_{{\rm II},1}}$, and the order $a_s^0$ of $\ln {Q_{{\rm I},2}}$ and $\ln {Q_{{\rm II},2}}$ are also equal. This shows a non-trivial equivalence of the two scale-setting methods at the non-conformal level, which means that PMC-I and PMC-II are the same at the NNLO level. Moreover, if we know additional higher-loop contributions, we can achieve more precise and closer PMC-I and PMC-II scales, and thus obtain smaller $\Delta_i$.

\begin{table}[htb]
\begin{tabular}{|c|c|c|c|}
\hline
           & ~~~$\Delta_{1}$~~~ & ~~~$\Delta_{2}$~~~ & ~~~$\Delta_{3}$~~~   \\
\hline
~~$R_{e^+e^-}(Q=31.6~{\rm GeV})$~~  & $-0.0043$ & $+0.0973$ & $+1.9389$  \\
\hline
~~$\Gamma^{\rm NS}(Z\to {\rm hadrons})$~~ & $-0.0030$ & $-0.0826$ & $+2.0432$ \\
\hline
~~$\Gamma \left( {\Upsilon (1S) \to {e^ + }{e^ - }} \right)$~~ & $+0.0353$ & $+0.1047$ & $+0.0816$ \\
\hline
~~$\Gamma(H\to b\bar{b})$~~ & $+0.0001$ & $-0.0014$ & $-0.0018$ \\
\hline
\end{tabular}
\caption{The logarithmic scale difference $\Delta_{i}$ for $R_{e^+ e^-}(Q=31.6{\rm GeV})$, $\Gamma^{\rm NS}(Z\to {\rm hadrons})$, $\Gamma \left( \Upsilon(1S) \to {e^+}{e^-} \right)$, and $\Gamma(H\to b\bar{b})$ up to four-loop QCD corrections, where ${\rm NS}$ stands for the non-singlet contribution.} \label{scalediff}
\end{table}

Table~\ref{scalediff} shows several four-loop examples of how $\Delta_i$ changes with the increment of the loop corrections. The four-loop expressions using conventional scale setting are adopted from Refs.\cite{Baikov:2005rw, Baikov:2008jh, Beneke:2007gj, Beneke:2005hg, Penin:2005eu, Beneke:2007pj, Beneke:2008cr, Baikov:2012er}. The PMC predictions for those channels can be found in Refs.\cite{Wang:2014aqa, Wang:2013bla, Wu:2014iba, Shen:2015cta}. In the case of $H\to b\bar{b}$, the computed scale differences are very small at any order. For example, the largest difference for the case of $H\to b\bar{b}$ is found between $Q_{{\rm I},3}$ and $Q_{{\rm II},3}$, which is less than $0.2\%$. The logarithmic scale differences for $R(e+e-)$ and $\Gamma^{\rm NS}(Z\to {\rm hadrons})$ are somewhat larger; their magnitudes follow the trend $|\Delta_1|\ll |\Delta_2| \ll |\Delta_3|$, indicating that the PMC-I and PMC-II scale differences quickly diminish as we include more $\{\beta_i\}$-terms to fix the PMC scales. In the case of $\Gamma \left( {\Upsilon (1S) \to {e^ + }{e^ - }} \right)$, the logarithmic scale differences $|\Delta_1|\ll |\Delta_2| \sim |\Delta_3|$ show that the scale differences between $Q_{{\rm I},3}$ and $Q_{{\rm II},3}$ are accidentally small even with less $\beta$-term information than the case of $Q_{{\rm I/II},2}$.

In the following, we will consider two observables $R_{e^+ e^-}(Q=31.6{\rm GeV})$ and $\Gamma(H\to b\bar{b})$ in detail to illustrate the differences and common features of the PMC-I and PMC-II predictions.

We rewrite the $e^+e^-$ annihilation $R$-ratio and the \mbox{$H \to b\bar{b}$} decay width as
\begin{eqnarray}
{R_{{e^ + }{e^ - }}}(Q) &=& 3\sum\limits_q {e_q^2} \left(1 +R_n^e(Q,{\mu})\right), \\
\Gamma (H \to b\bar{b}) &=& \frac{{3{G_F}{M_H}m_b^2}}{{4\sqrt 2 \pi }}\left(1 + R_n^H(Q,\mu)\right),
\end{eqnarray}
where $R_n^{e/H}(Q,{\mu}) = \sum\nolimits_{i = 0}^n {\cal C}_i^{e/H} (Q,\mu)a_s^{i + 1}(\mu)$ and the scale $\mu$ is arbitrary. Here $Q$ stands for a typical momentum flow of the process. In order to compare with data, we will take $Q=31.6$ GeV for the $R$-ratio and $Q=M_H$ for $H\to b\bar{b}$. The coefficients ${\cal C}_i^{e/H}$ with its explicit $n_f$-dependence up to four-loop level can be found in Refs.~\cite{Baikov:2005rw,Baikov:2008jh}. We adopt four-loop $\alpha_s$-running and fix the QCD parameter $\Lambda_{\overline {\rm MS}}$ by using $\alpha_s({M_z}) = 0.1185 \pm 0.0006$~\cite{Agashe:2014kda}.

\begin{table}[htb]
\begin{tabular}{cccccc}
\hline
           &~~~LO~~~  &~~NLO~~ &$~\rm{N^2LO}~$& $~\rm{N^3LO}~$ &~ \rm{total} ~ \\
\hline
PMC-I    & $0.04294$ & $0.00340$ & $-0.00002$ & $-0.00001$ & $0.04631$ \\
PMC-II   & $0.04290$ & $0.00352$ & $-0.00004$ & $-0.00002$ & $0.04636$ \\
Conv.    & $0.04499$ & $0.00285$ & $-0.00117$ & $-0.00033$ & $0.04634$ \\
\hline
\end{tabular}
\caption{The contributions of each of the contributions ($\rm{LO}$, $\rm{NLO}$, $\rm{N^2LO}$ and $\rm{ N^3LO}$) to the four-loop pQCD approximate $R_3^e$.} \label{eeP}
\end{table}

\begin{table}[htb]
\begin{tabular}{cccccc}
\hline
           &~~~LO~~~  &~~NLO~~ &$~\rm{N^2LO}~$& $~\rm{N^3LO}~$ &~ \rm{total} ~  \\
\hline
PMC-I    & $0.2268$ & $0.0249$ & $-0.0091$ & $-0.0012$ & $0.2414$ \\
PMC-II   & $0.2268$ & $0.0249$ & $-0.0094$ & $-0.0012$ & $0.2411$ \\
Conv.    & $0.2037$ & $0.0377$ & $+0.0019$ & $-0.0014$ & $0.2419$ \\
\hline
\end{tabular}
\caption{The contributions of each of the contributions ($\rm{LO}$, $\rm{NLO}$, $\rm{N^2LO}$ and $\rm{ N^3LO}$) to the four-loop pQCD approximate $R_3^H$.} \label{hhP}
\end{table}

Tables~\ref{eeP} and \ref{hhP} show the contributions of each contribution up to the four-loop pQCD term $R^{e/H}_3$. The pQCD predictions using conventional scale setting with $\mu=Q$ are also presented for comparison. We have set the N$^3$LO scale $Q_{{\rm I},4}$ ($Q_{{\rm II},4}$) to the be highest order determined scale $Q_{{\rm I},3}$ ($Q_{{\rm II},3}$), following the prescription of PMC-I. Tables~\ref{eeP} and \ref{hhP} show that the pQCD convergence for the PMC-I and PMC-II are very similar, which are, as required, better than the conventional one due to the elimination of divergent renormalon terms. The pQCD series for $H\to b\bar{b}$ is almost the same. There are slight differences for $R(e^+ e^-)$ due to the logarithmic scale differences listed in Table~\ref{scalediff}; however, these are greatly suppressed by the magnitudes of the conformal coefficients and the $\alpha_s$-powers.

\begin{figure}[tb]
\includegraphics[width=0.50\textwidth]{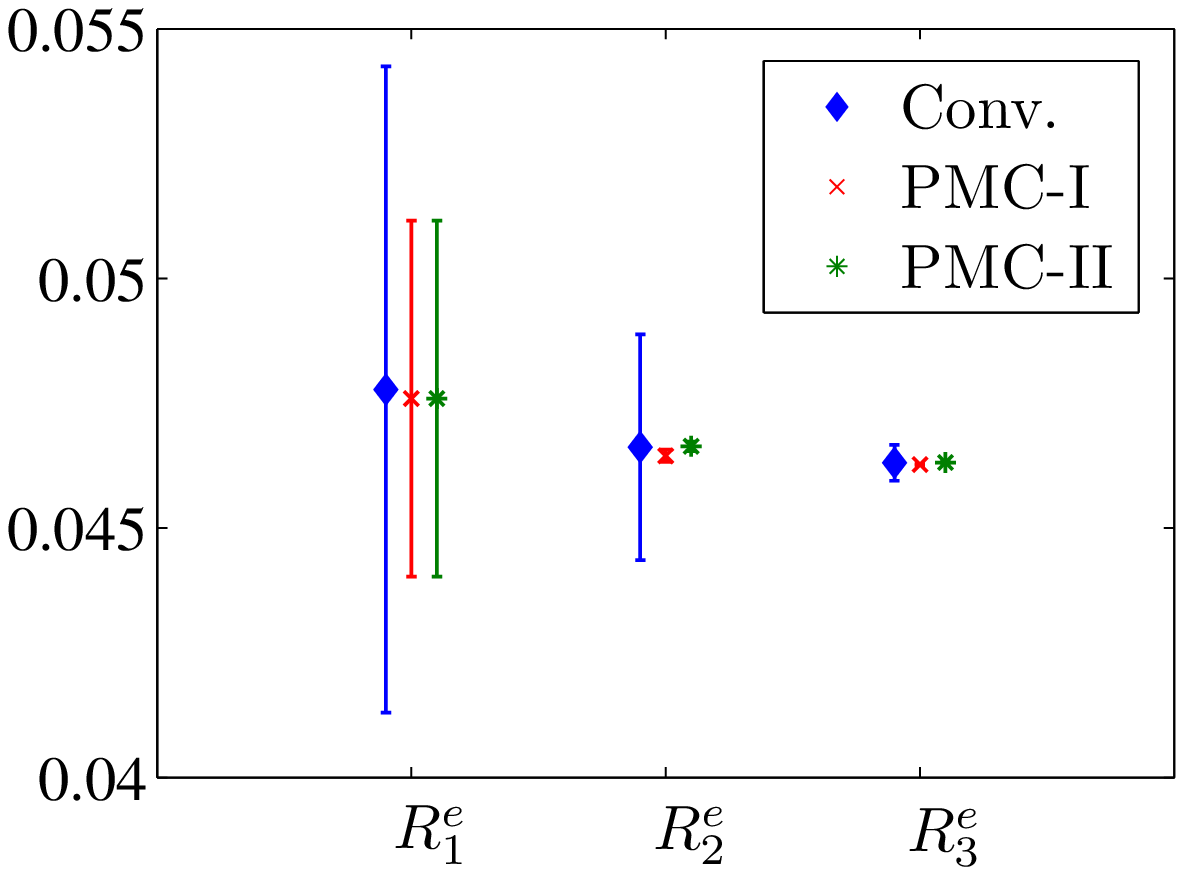}
\includegraphics[width=0.50\textwidth]{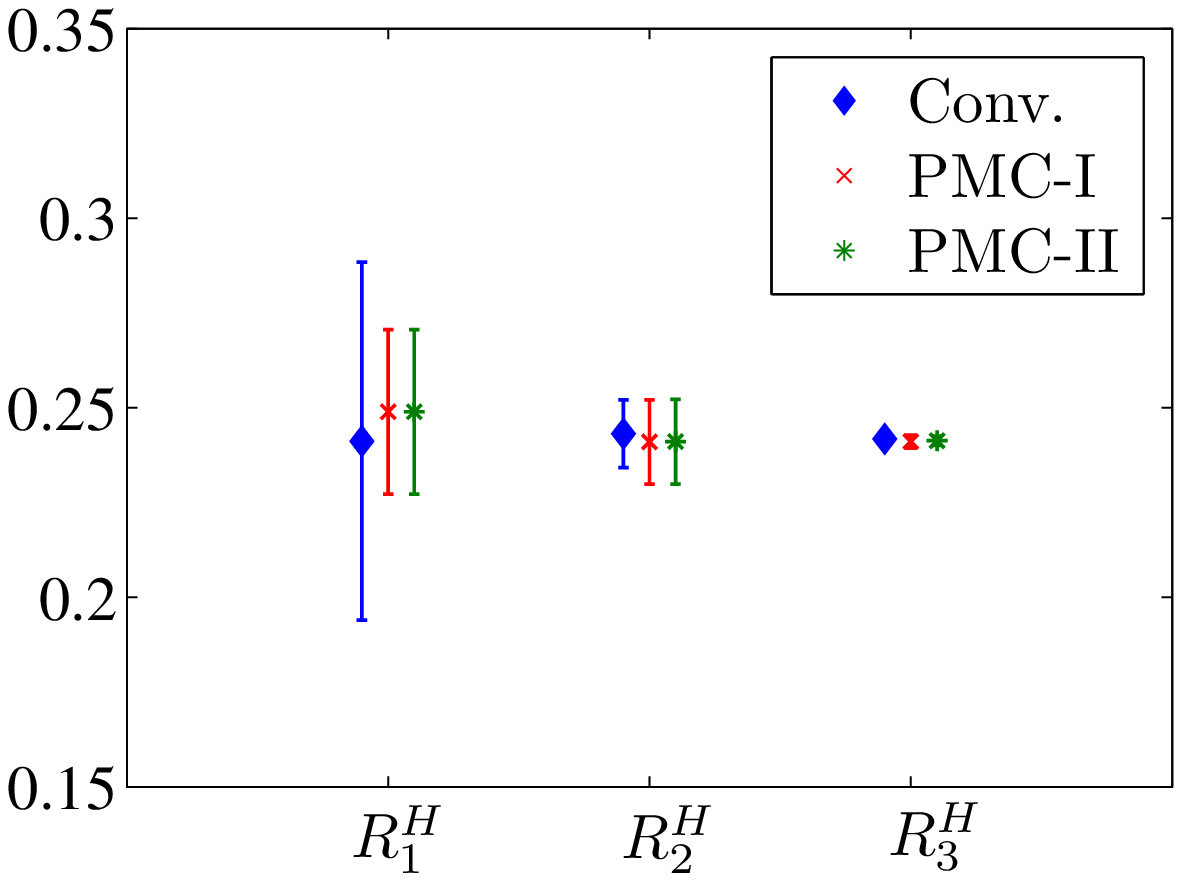}
\caption{Results for $R_n^{e}$ and $R_n^{H}$ up to $(n+1)$-loop QCD corrections together with their approximate errors $\pm |\tilde{\cal C}_n^{e/H} a_s^{n + 1}|_{\rm MAX}$ under the PMC-I and PMC-II methods, where ${\tilde{\cal C}}_n^{e/H}$ is the $(n+1)_{\rm th}$-order conformal coefficient and $n=(1,2,3)$. } \label{Plot}
\end{figure}

FIG.~\ref{Plot} presents the results for $R_n^{e/H}$ up to $(n+1)$-loop QCD corrections, together with their estimated errors $\pm |\tilde{\cal C}_n^{e/H}a_s^{n + 1}|_{\rm MAX}$ using both the PMC-I and PMC-II approaches. Here, ${\tilde{\cal C}}_n^{e/H}$ are $(n+1)_{\rm th}$-order conformal coefficients, and the subscript ``MAX" means the maximum value of $|\tilde{\cal C}_n^{e/H}a_s^{n + 1}|$ obtained by varying the scale $\mu$ within the usual region of $[Q/2,2Q]$. This error estimate is natural for the PMC, since after applying the PMC-I or PMC-II, the pQCD convergence is ensured and the only uncertainty is from the last term of the pQCD series due to the unfixed PMC scale at this particular order. When additional pQCD loop corrections are taken into consideration, one obtains a weaker scale dependence. This agrees with the conventional wisdom that as one incorporates a higher order calculation, one can obtain an increasingly reliable scale-invariant estimate. The PMC-I and PMC-II predictions are very close in magnitude; their values quickly approach convergence, indicating that a low-order calculation could be sufficient to achieve an accurate pQCD prediction.

Let us end with a final comment on the generality of the degeneracy relations found using the ${\cal R}_\delta$-scheme~\cite{Mojaza:2012mf}. The degeneracy relations actually hold under any scheme, which was pointed out in Ref.~\cite{Brodsky:2013vpa}. To explain this, we adopt the effective charge method introduced by Grunberg~\cite{Grunberg:1982fw}. Any effective charge $a_A$ of a physical observable $A$ can be expanded over the running coupling $a_{\cal R}$ of the ${\cal R}_\delta$-scheme as
\begin{eqnarray}
a_A(\mu) &=& a_{\cal R}(\mu) + [r^A_{2,0} + \beta_0 r^A_{2,1} ] a_{\cal R}^2(\mu) + [r^A_{3,0} + \nonumber\\
&& \beta_1 r^A_{2,1} + 2 \beta_0 r^A_{3,1} + \beta_0^2 r^A_{3,2} ] a^3_{\cal R}(\mu) + \nonumber\\
&&[r^A_{4,0} + \beta_2^{\cal R} r^A_{2,1}  + 2 \beta_1 r^A_{3,1} +  \frac{5}{2} \beta_1 \beta_0 r^A_{3,2} + \nonumber\\
&& 3 \beta_0 r^A_{4,1} + 3 \beta_0^2 r^A_{4,2} + \beta_0^3 r^A_{4,3} ] a_{\cal R}(\mu)^4  +\cdots,
\end{eqnarray}
where the universality of $\beta_0$ and $\beta_1$ is used, and the explicit scheme-dependence of $\beta_2$ is expressed. The $\beta$-function of the $A$-scheme is related to the one of the $R_\delta$-scheme through the identity, $\beta^A(a_A) = \frac{{\partial a_A}}{{\partial a_{\cal R}}}{\beta^{\cal R}}({a_{\cal R}})$. If one analyzes another effective charge $a_B$ in the same way, one finds that $a_A$ can also be expanded over $a_B$, through scheme-transformations mediated by the $R_\delta$-scheme
\begin{eqnarray}
{a_A}(\mu) &=& {a_B}(\mu) + (r_{2,0}^{AB} + \beta_0 r_{2,1}^{AB})a_B^2(\mu)+ (r_{3,0}^{AB} + \nonumber\\
&& \beta_1 r_{2,1}^{AB} + 2\beta_0 r_{3,1}^{AB} + \beta_0^2 r_{3,2}^{AB}) a_B^3(\mu) + \nonumber\\
&& (r_{4,0}^{AB} + \beta_2^A r_{2,1}^{AB} + 2\beta_1 r_{3,1}^{AB} + \frac{5}{2} \beta_0 \beta_1 r_{3,2}^{AB} + \nonumber\\
&& 3\beta_0 r_{4,1}^{AB} + 3\beta_0^2 r_{4,2}^{AB} + \beta_0^3 r_{4,3}^{AB}) a_B^4(\mu) +\cdots.
\end{eqnarray}
This shows that the degeneracy relations still hold, even if scheme $B$ is not an MS-like scheme. The coefficients $r_{i,j}^{AB}$ up to four-loop level can be found in Ref.~\cite{Brodsky:2013vpa}~\footnote{There is a typo in Eq.(93i), i.e. the last term $(-3r^{B}_{3,2}r^{AB}_{2,0})$ should be corrected as $(-3r^{B}_{3,2}r^{AB}_{2,1})$}. Among them the conformal ones are
\begin{eqnarray}
r_{2,0}^{AB} &=& r_{2,0}^A - r_{2,0}^B, \\
r_{3,0}^{AB} &=& r_{3,0}^A - r_{3,0}^B - 2r_{2,0}^B r_{2,0}^{AB}, \\
r_{4,0}^{AB} &=& r_{4,0}^A - r_{4,0}^B - 3r_{2,0}^B r_{3,0}^{AB} - (r_{2,0}^{{B^2}} + 2r_{3,0}^B) r_{2,0}^{AB},
\end{eqnarray}
which are purely expressed in terms of the conformal coefficients $r_{i,0}^{A/B}$ in the ${\cal R}_\delta$-scheme. This self-consistency shows that the applicability of the PMC is ensured and its predictions are scheme-independent.

In conclusion, we have shown that the two all-orders PMC approaches are equivalent to each other at the level of conformality and are thus equally viable PMC procedures. PMC-I implementation is a direct extension of the BLM approach, whereas PMC-II provides additional theoretical improvements; in addition, it can be readily automatized using the $R_\delta$-scheme. By construction, both the PMC-I and PMC-II satisfy all of the principles of the renormalization group, thus providing scale-fixed and scheme-independent predictions at any fixed order. Those two implementations of PMC differ, however, at the non-conformal level, by predicting slightly different RG scales of the running coupling. This difference arises due to different ways of resumming the non-conformal terms, but this difference decreases rapidly when additional loop corrections are included.

The key step of PMC-II is to use the pattern generated by the RG-equation and its degeneracy relations to identify which terms in the pQCD series are associated with the QCD $\beta$-function and which terms remain in the $\beta=0$ conformal limit. The $\beta$-terms are then systematically absorbed by shifting the scale of the running coupling at each order, thus providing the PMC scheme-independent prediction. The recursive patterns and degeneracy relations between the $\beta$-terms at each order are essential for carrying out this procedure. The implementation of PMC-II illuminates how the renormalization scheme and initial scale dependence are eliminated at each order. These advantages shows the PMC-II is theoretically robust and is the preferred method for practical implementations of PMC. \\

\noindent{\bf Acknowledgements:} This work was supported in part by the Natural Science Foundation of China under Grant No.11275280, the Fundamental Research Funds for the Central Universities under Grant No.CDJZR305513, and the Department of Energy Contract No.DE-AC02-76SF00515. SLAC-PUB-16287. Nordita-2015-54.

\end{document}